\documentclass[conference, 11pt]{IEEEtran}
\IEEEoverridecommandlockouts
\usepackage{cite}
\usepackage{amsmath,amssymb,amsfonts}
\usepackage{algorithmic}
\usepackage{graphicx}
\usepackage{textcomp}
\usepackage{xcolor}

\DeclareMathAlphabet{\mathsfbr}{OT1}{cmss}{m}{n}
\SetMathAlphabet{\mathsfbr}{bold}{OT1}{cmss}{bx}{n}
\DeclareRobustCommand{\msf}[1]{%
  \ifcat\noexpand#1\relax\msfgreek{#1}\else\mathsfbr{#1}\fi
}

\makeatletter
\newcommand{\msfgreek}[1]{\csname s\expandafter\@gobble\string#1\endcsname}
\makeatother

\DeclareFontEncoding{LGR}{}{} 
\DeclareSymbolFont{sfgreek}{LGR}{cmss}{m}{n}
\SetSymbolFont{sfgreek}{bold}{LGR}{cmss}{bx}{n}
\DeclareMathSymbol{\salpha}{\mathord}{sfgreek}{`a}
\DeclareMathSymbol{\sbeta}{\mathord}{sfgreek}{`b}
\DeclareMathSymbol{\sgamma}{\mathord}{sfgreek}{`g}
\DeclareMathSymbol{\sdelta}{\mathord}{sfgreek}{`d}
\DeclareMathSymbol{\sepsilon}{\mathord}{sfgreek}{`e}
\DeclareMathSymbol{\szeta}{\mathord}{sfgreek}{`z}
\DeclareMathSymbol{\seta}{\mathord}{sfgreek}{`h}
\DeclareMathSymbol{\stheta}{\mathord}{sfgreek}{`j}
\DeclareMathSymbol{\siota}{\mathord}{sfgreek}{`i}
\DeclareMathSymbol{\skappa}{\mathord}{sfgreek}{`k}
\DeclareMathSymbol{\slambda}{\mathord}{sfgreek}{`l}
\DeclareMathSymbol{\smu}{\mathord}{sfgreek}{`m}
\DeclareMathSymbol{\snu}{\mathord}{sfgreek}{`n}
\DeclareMathSymbol{\sxi}{\mathord}{sfgreek}{`x}
\DeclareMathSymbol{\somicron}{\mathord}{sfgreek}{`o}
\DeclareMathSymbol{\spi}{\mathord}{sfgreek}{`p}
\DeclareMathSymbol{\srho}{\mathord}{sfgreek}{`r}
\DeclareMathSymbol{\ssigma}{\mathord}{sfgreek}{`s}
\DeclareMathSymbol{\stau}{\mathord}{sfgreek}{`t}
\DeclareMathSymbol{\supsilon}{\mathord}{sfgreek}{`u}
\DeclareMathSymbol{\sphi}{\mathord}{sfgreek}{`f}
\DeclareMathSymbol{\schi}{\mathord}{sfgreek}{`q}
\DeclareMathSymbol{\spsi}{\mathord}{sfgreek}{`y}
\DeclareMathSymbol{\somega}{\mathord}{sfgreek}{`w}

\DeclareMathSymbol{\svarsigma}{\mathord}{sfgreek}{`c}

\DeclareMathSymbol{\sGamma}{\mathalpha}{sfgreek}{`G}
\DeclareMathSymbol{\sDelta}{\mathalpha}{sfgreek}{`D}
\DeclareMathSymbol{\sTheta}{\mathalpha}{sfgreek}{`J}
\DeclareMathSymbol{\sLambda}{\mathalpha}{sfgreek}{`L}
\DeclareMathSymbol{\sXi}{\mathalpha}{sfgreek}{`X}
\DeclareMathSymbol{\sPi}{\mathalpha}{sfgreek}{`P}
\DeclareMathSymbol{\sSigma}{\mathalpha}{sfgreek}{`S}
\DeclareMathSymbol{\sUpsilon}{\mathalpha}{sfgreek}{`U}
\DeclareMathSymbol{\sPhi}{\mathalpha}{sfgreek}{`F}
\DeclareMathSymbol{\sPsi}{\mathalpha}{sfgreek}{`Y}
\DeclareMathSymbol{\sOmega}{\mathalpha}{sfgreek}{`W}

\DeclareRobustCommand{\mcal}[1]{%
  \ifcat\noexpand#1\relax\mathnormal{#1}\else\cal{#1}\fi
}
\DeclareRobustCommand{\BM}[1]{%
  \ifcat\noexpand#1\relax\bm{\boldUppercaseItalicGreek{#1}}\else\bm{#1}\fi
}
\makeatletter
\newcommand{\boldUppercaseItalicGreek}[1]{\csname var\expandafter\@gobble\string#1\endcsname}
\makeatother

\usepackage{bm}
\usepackage{color, courier}
\usepackage{psfrag}
\usepackage{acronym}
\usepackage{amsmath}
\usepackage{amssymb}
\usepackage{amsfonts}
\usepackage[colorlinks = true, allcolors = black]{hyperref}
\usepackage{latexsym}
\usepackage{mathrsfs}
\usepackage{epstopdf}
\usepackage{algorithm}
\usepackage{mathtools}
\usepackage[caption=false, font=scriptsize]{subfig}
\usepackage{algorithmic}
\usepackage{relsize}
\usepackage{comment}

\graphicspath{{figures/}}

\newcommand{\st}{\operatorname{s.t.}\,}

\definecolor{green}{rgb}{0, 0.5, 0}
\definecolor{pink}{rgb}{1, 0, 1}

\definecolor{bg}{RGB}{199, 237, 204}

\acrodef{agi}[AgI]{augmented information}
\acrodef{ldp}[LDP]{Lyapunov drift-plus-penalty}
\acrodef{lp}[LP]{linear programming}
\acrodef{ap}[AP]{access point}
\acrodef{pdf}[pdf]{probability density function}
\acrodef{ue}[UE]{user equipment}
\acrodef{sfc}[SFC]{service function chain}
\acrodef{bp}[BP]{back-pressure}
\acrodef{sinr}[SINR]{signal-to-interference-plus-noise ratio}
\acrodef{mec}[MEC]{mobile edge computing}
\acrodef{csi}[CSI]{channel state information}

\acrodef{nfv}[NFV]{network function virtualization}
\acrodef{sdn}[SDN]{software defined networking}

\acrodef{wrt}[w.r.t.]{with respect to}
\acrodef{wlog}[w.l.o.g.]{Without loss of generality}

\acrodef{umw}[UMW]{universal max-weight}

\def\BibTeX{{\rm B\kern-.05em{\sc i\kern-.025em b}\kern-.08em
    T\kern-.1667em\lower.7ex\hbox{E}\kern-.125emX}}
    
\begin{document}

\title{Compute- and Data-Intensive Networks:  \\
The Key to the Metaverse
\thanks{Prof. Antonia M. Tulino is also with University\`{a} degli Studi di Napoli Federico II, Naples 80138, Italy.
\newline \indent This work was supported by the National Science Foundation (NSF) under CNS-1816699.}
}

\author{\IEEEauthorblockN{Yang Cai}
\IEEEauthorblockA{
\textit{\small University of Southern California}\\
\small Los Angeles, USA \\
yangcai@usc.edu}
\and
\IEEEauthorblockN{Jaime Llorca}
\IEEEauthorblockA{
\textit{\small New York University}\\
\small New York City, USA \\
jllorca@nyu.edu}
\and
\IEEEauthorblockN{Antonia M. Tulino}
\IEEEauthorblockA{
\textit{\small New York University}\\
\small New York City, USA \\
atulino@nyu.edu}
\and
\IEEEauthorblockN{Andreas F. Molisch}
\IEEEauthorblockA{
\textit{\small University of Southern California}\\
\small Los Angeles, USA \\
molisch@usc.edu}
}

\maketitle

\IEEEpubid{
\begin{minipage}{2\columnwidth}
    \centering
    {\footnotesize
    \vspace{80pt}
    This work has been submitted to the {IEEE} for possible publication. Copyright may be transferred without notice, after which this version may no longer be accessible.
    }
\end{minipage}
}

\begin{abstract}

The worlds of computing, communication, and storage have for a long time been treated separately, and even the recent trends of cloud computing, distributed computing, and mobile edge computing have not fundamentally changed the role of networks, still designed to move data between end users and pre-determined computation nodes, without true optimization of the end-to-end compute-communication process. However, the emergence of Metaverse applications, where users consume multimedia experiences that result from the real-time combination of distributed live sources and stored digital assets, has changed the requirements for, and possibilities of, systems that provide distributed caching, computation, and communication. We argue that the real-time interactive nature and high demands on data storage, streaming rates, and processing power of Metaverse applications will accelerate the merging of the cloud into the network, leading to highly-distributed tightly-integrated compute- and data-intensive networks becoming universal compute platforms for next-generation digital experiences. In this paper, we first describe the requirements of Metaverse applications and associated supporting infrastructure, including relevant use cases. We then outline a comprehensive {\em cloud network flow} mathematical framework, designed for the end-to-end optimization and control of such systems, and show numerical results illustrating its promising role for the efficient operation of Metaverse-ready networks.
\end{abstract}


\section{Introduction}

Next-generation (NextG) networks are rapidly evolving towards tightly integrated computing, caching and communication (3C) systems that go beyond current
(i) computation-centric cloud data centers interconnected by a wide area network,
(ii) communication-centric 5G networks connecting mobile users to cloud resources, and
(iii) caching-centric content distribution networks.
NextG networks are envisioned to be highly distributed mobile/wireless-first 3C systems, where a wide range of distributed network elements, including end devices, access points, and edge/cloud servers cooperate contributing resources and participating in the routing, storage, and processing functions needed to deliver the services that will define the future of consumer experiences and industrial automation. 

Indeed, the ubiquity of live and stored data sources (e.g., real-world sensors and digital assets) fueled by the convergence of physical reality and digital virtuality, and the seamless access to distributed computational resources enabled by NextG networks will essentially blur the space- and time-scale separation between data collection, information processing, and experience delivery, enabling a new breed of {\em Metaverse applications} that will transform the way we live, work, and interact with the physical world. It is envisioned that Metaverse applications will drive massive investments toward the digitization, automation, and enhanced interactivity of physical systems and human experiences. Augmented/virtual/extended reality (XRs), telepresence, immersive video, digital twins, and multi-player gaming are all examples of Metaverse applications that require real-time (i) aggregation of distributed data streams, (ii) in-network data processing and information synthesis, and (iii) distribution of highly personalized streams to multiple interacting users.

We argue that the unprecedented communication, computation, and storage requirements imposed by the Metaverse will demand a new universal compute platform driven by the efficient integration of 3C technologies into NextG networks, along with new tools and methods for the end-to-end optimization and dynamic control of the resulting global infrastructure. To this end, in this paper:
\begin{itemize}
    \item We illustrate the relevance and generality of Metaverse applications, as well as their unique multi-dimensional requirements.
    \item We describe the main characteristics of the envisioned Metaverse-ready compute- and data-intensive infrastructure.
    \item We outline a cloud network flow mathematical framework, especially suited for the integrated control of 3C networks and the end-to-end optimization of Metaverse application delivery.
    \item We show numerical results that validate the promising role of the described framework to enable the efficient operation of Metaverse-ready networks.
\end{itemize}

\section{Metaverse Applications}
\label{sec:applications}

The Metaverse, etymologically a combination of {\em meta} (i.e., {\em beyond}) and {\em universe}, refers to a computer-generated world that flexibly blends physical reality and digital virtuality in order to provide immersive, interactive, realistic, and augmented digital experiences, with a wide spectrum of consumer and industrial applications.

{\bf Social:}
The Metaverse, which aggregates and transcends traditional media (text, audio, image, video), will shape the future of online social networks, initially built to connect individual users for communication and content sharing. For example, Meta\textsuperscript{\textregistered} (formerly known as Facebook\textsuperscript{\textregistered}) is actively developing social VR platforms where users can have digital representations (avatars), interact with each other, and participate in shared activities like shopping, tourism, watching movies, and attending events.

{\bf Gaming:}
The Metaverse has also entered the gaming space (e.g., VR/AR games), making games more realistic and interactive by accelerating efforts on physical world digitization and the enablement of multisensory experiences. Metaverse games such as Horizon Worlds\textsuperscript{\textregistered}, Roblox\textsuperscript{\textregistered}, and Fortnite\textsuperscript{\textregistered} have become widely popular, and movies like Ready Player One\textsuperscript{\textregistered} and Free Guy\textsuperscript{\textregistered} envision a more interactive mixed-reality world in future Metaverse games.

\begin{figure}[t]
    \centering
    \subfloat[Social]{
    \includegraphics[width = 0.45 \columnwidth]{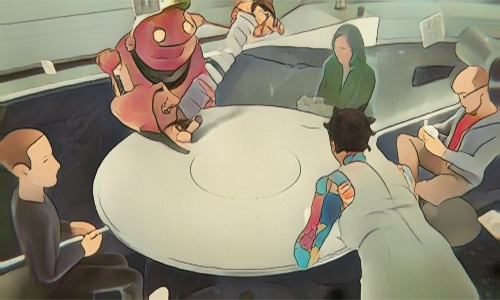}
    }\hfill
    \subfloat[Gaming]{
    \includegraphics[width = 0.45 \columnwidth]{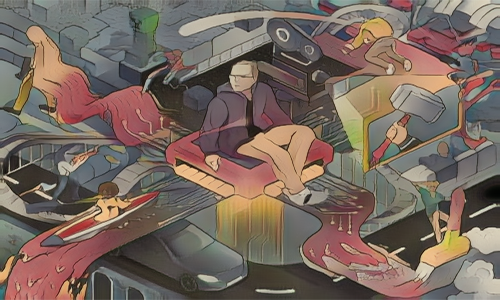}
    }
    
    \subfloat[Industry]{
    \includegraphics[width = 0.45 \columnwidth]{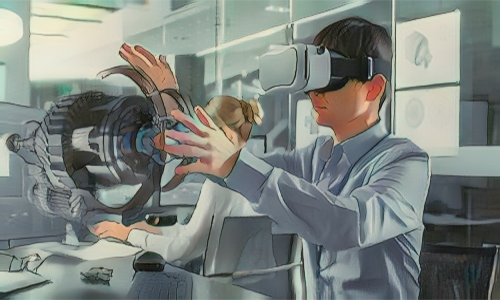}
    }\hfill
    \subfloat[Collaboration]{
    \includegraphics[width = 0.45 \columnwidth]{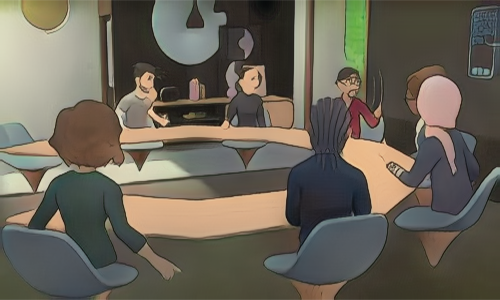}
    }
    
    \subfloat[Health]{
    \includegraphics[width = 0.45 \columnwidth]{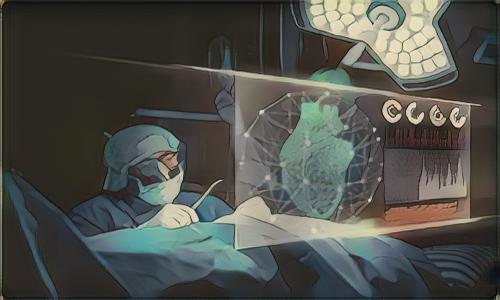}
    }\hfill
    \subfloat[Education]{
    \includegraphics[width = 0.45 \columnwidth]{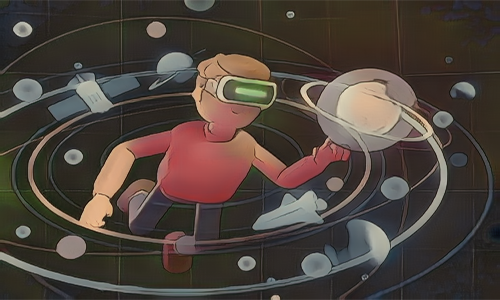}
    }
    \caption{Metaverse applications.}
    \label{fig:use_cases}
\end{figure}

{\bf Industry:}
In industrial applications, the Metaverse can increase productivity along multiple phases of a product's life cycle. First, product design can be conducted in the Metaverse running accurate simulations, at a lower cost and a faster pace than when creating physical design samples. Second, the Metaverse can use digital twins to enhance operational efficiency and reduce quality control risks in the manufacturing process. Finally, the Metaverse provides a communal space where interdisciplinary teams and customers can have real-time interactions, increasing the efficiency and agility of the product design and development life cycle. The NVIDIA\textsuperscript{\textregistered} Omniverse\textsuperscript{\texttrademark} Platform is a clear example of a relevant tool for industrial Metaverse applications.

{\bf Collaboration:}
In the Metaverse, people can create a personalized workspace or virtual office -- a more flexible working environment to facilitate collaboration regardless of geographical restrictions. Digital assets representing people (e.g., avatars) and objects can be incorporated as needed into the 3D space, giving a new dimension to today's online meetings.

{\bf Health:}
The Metaverse also enables important applications in the healthcare industry, including tele-medicine, augmented fitness, and in particular, remote surgery. When provided highly realistic environments, remote doctors can perform high-precision operations on a patient's body. Besides, real-time physical conditions of the patient can enrich displayed contents to help doctors' decision making.

{\bf Education:}
The Metaverse transforms the way knowledge is presented. In descriptive or explanatory courses, students can be exposed to visual 3D models with improved clarity compared to any precedent media. In training courses, students can practice their skills in realistic environments and enjoy efficient, low-risk learning experiences. For example, it enables learning how to manipulate hazardous substances without actual exposure to those substances.

\section{Metaverse Infrastructure}

In this section, we summarize the unprecedented resource requirements imposed by Metaverse applications, and then describe the main characteristics of the envisioned supporting infrastructure.

\subsection{Resource Requirements}
\label{sec:requirements}

As illustrated in Section \ref{sec:applications}, the Metaverse is foreseen to become a global digital platform where users can consume real-time interactive experiences that seamlessly blend physical reality and digital virtuality. Going beyond content retrieval-and-distribution (e.g., web, video streaming), in the Metaverse, user experiences result from the real-time aggregation, processing/composition, and delivery of multiple live streams and digital assets.

In the following, we describe the computation, storage and communication requirements of Metaverse applications, illustrated in the context of a VR streaming application in Fig. \ref{fig:requirement}.

\subsubsection{Computation Requirements}

Central to Metaverse applications is the blending of physical and digital worlds into rich multimedia immersive environments -- a task of high computational demand. 

In industrial Metaverse applications, massive computational resources are consumed to build physically accurate simulation environments. Prospective consumer applications will also challenge computing power requirements. For example, running a typical AAA game (e.g., Fortnite\textsuperscript{\textregistered}) today requires multiple teraFLOPS of graphics horsepower, and the demand is expected to grow by two orders of magnitude to create fully immersive Metaverse experiences \cite{credit-suisse2022metaverse}.
Even basic video processing tasks (e.g., object recognition and tracking) can exhibit substantial computational complexity when applied to increasingly enriched virtual worlds involving massive numbers of users and digital assets.

In addition, Metaverse applications may involve multiple processing tasks (running as separate service functions) operating on source and intermediate data streams for the generation of the consumable experiences (see, e.g., coding, decoding, and rendering functions for VR streaming in Fig. \ref{fig:requirement}). The requirements of each service function may also depend on the hosting hardware, e.g, general-purpose Central Processing Units (CPUs), Graphic Processing Units (GPUs), or tailor-made computing hardware.

\subsubsection{Storage Requirements}
\label{sec:data}

An explosively growing number of digital assets (representing objects, spaces, attributes, value, etc.) are crowding into the Metaverse.

\begin{figure}[t]
    \centering
    \includegraphics[width = .9 \columnwidth]{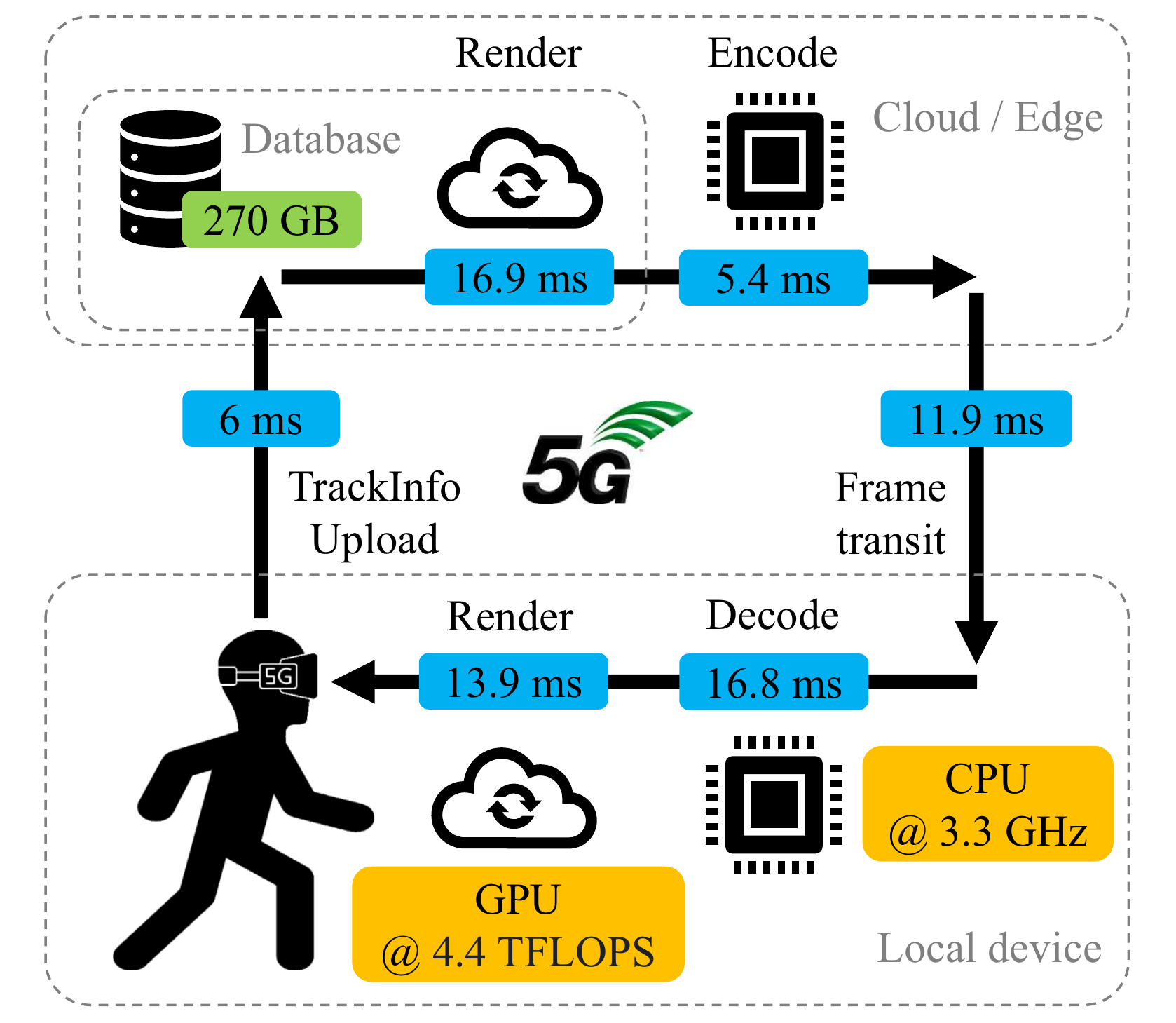}
    \caption{Resource consumption of {\em early-stage} VR streaming \cite{hu2020requirement}. The {\em computation} requirements are highlighted in yellow and the {\em storage} requirement for a $5$-minute $360^\circ$ video \cite{sun2019mobileVR} in green. The system attains an end-to-end delay (sum of individual delays in blue) of $70.9$ ms, and a perception delay (with motion prediction) of $8.2$ ms.}
    \label{fig:requirement}
\end{figure}

As key building blocks of Metaverse applications, digital assets are used to compose the immersive experiences consumed by users according to their real-time interactions. For example, digital twins representing the properties of physical systems are essential components of industrial Metaverse applications; another consumer application example, VR streaming, renders 3D scenarios from digital scenes (see Fig. \ref{fig:requirement}). These storage-hungry applications will impose tremendous resource requirements. For example, {\em Entry-level} VR, designated to support $8$K-resolution $360^\circ$ video streaming for $20$-minute experience duration, requires around $10$ TB
storage space for the produced uncompressed video. Such already high demand is expected to increase in future phases of {\em Advanced} VR (by a factor of 10) and {\em Ultimate} VR (by a factor of 100) \cite{hu2020requirement}.

\subsubsection{Communication Requirements}

The Metaverse is created to power interactions among the users and with the virtual environment, which requires the real-time aggregation of multiple live streams and digital assets, and the distribution of the resulting processed/composed streams.

Fueled by growing trends in wearable devices and the Internet of things (IoT), a massive number of sensors will continuously collect data about the physical world as inputs to the Metaverse. The real-time aggregation of the resulting live data streams, as well as pre-produced digital assets, will continue to accelerate network traffic growth.
As projected in \cite{credit-suisse2022metaverse}, the Metaverse could drive video traffic -- which already accounts for $80\%$ of today's Internet traffic and grows at a $30\%$ compound annual growth rate (CAGR) -- to grow at a $37\%$ CAGR, leading to $24$ times the current data usage over the next decade.

Even more critical is the end-to-end latency requirements. To enable real-time interaction, most Metaverse applications will impose end-to-end delay constraints of less than $20$ ms (e.g., $7$ to $15$ ms) \cite{sun2019mobileVR}, which could go as low as $1$ ms for tactile applications such as remote surgery. Exceeding these stringent latency requirements not only leads to lagged responses, but also causes discomfort, e.g., dizziness and nausea, in VR applications.

\subsection{Envisioned Infrastructure}
\label{sec:NextG_net}

\begin{figure}[t]
    \centering
    \includegraphics[width = .9 \columnwidth]{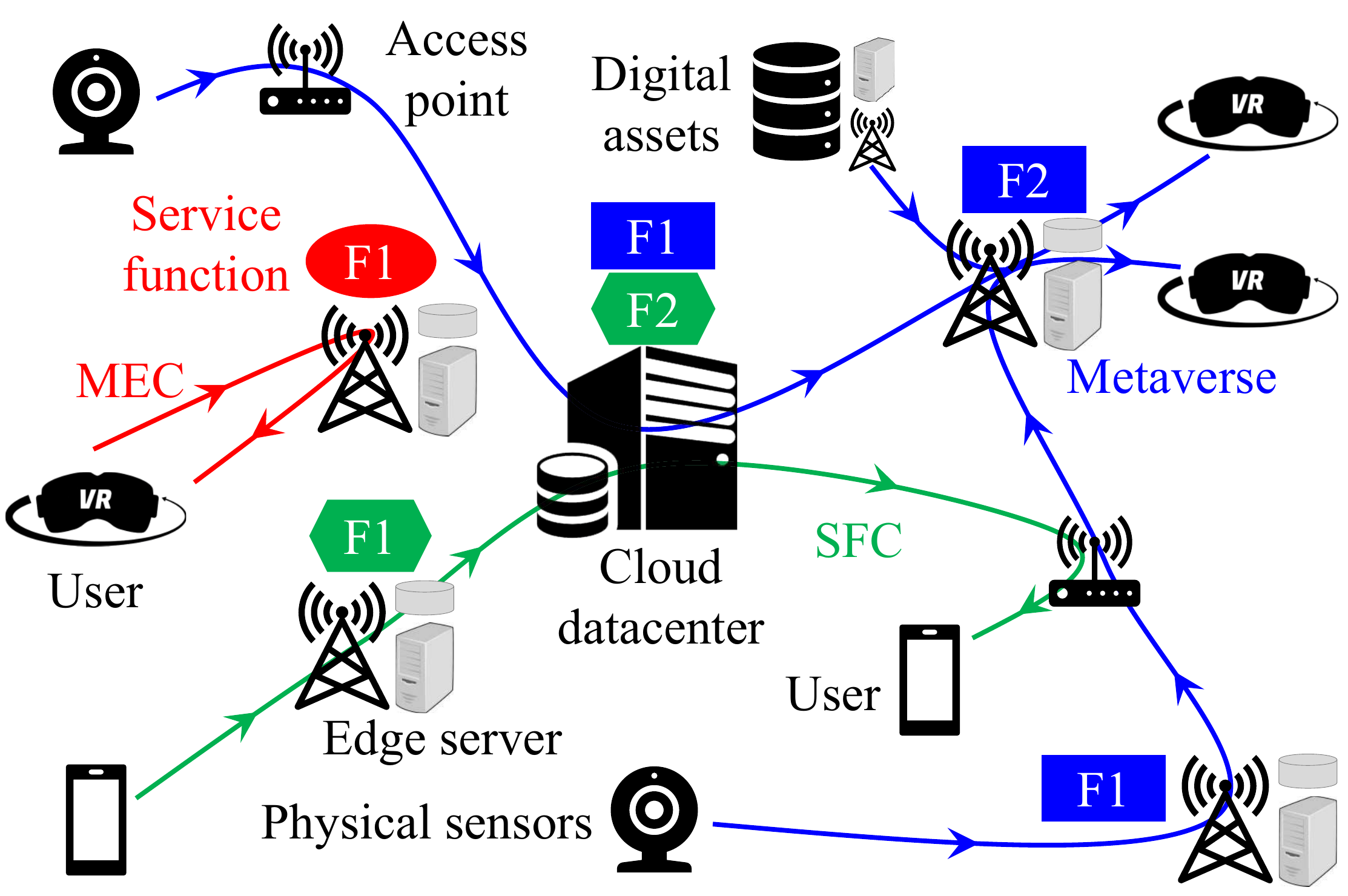}
    \caption{The envisioned infrastructure for the Metaverse, in which MEC and SFC services can be handled as special cases of general Metaverse applications (see also Fig. \ref{fig:service}).}
    \label{fig:CCC_network}
\end{figure}

While initial VR/AR applications are being supported by local compute platforms such as VR headsets and/or gaming consoles, the full breadth of Metaverse applications, involving distributed remote sources, dispersed users, and multiple service functions, imposing the level of resource requirements described in the previous section, will demand a much more powerful global-scale infrastructure. 

To this end, we argue that the Metaverse supporting infrastructure shall be a {\em universal compute platform running on a highly-distributed tightly-integrated compute-, communication- and data-intensive NextG network}. As shown in Fig. \ref{fig:CCC_network}, the envisioned NextG network interconnects computing- and caching-enabled user devices, access points, edge servers, and cloud data centers, along a device-edge-cloud continuum. Each node is able to host a set of service functions (depending on their capabilities) that can be dynamically activated to run the supported computation tasks; in addition, each node may use its available local storage to cache Metaverse digital assets. Each link is capable of data transmission between connected devices, laying the foundation for cooperative computing and caching. Aided by advanced network programmability (e.g., software defined networking, SDN) and virtualization (e.g., network function virtualization, NFV) technologies, service functions can be flexibly interconnected and elastically executed at different network locations, while efficiently accessing required digital assets cached throughout the network. 

While the worlds of computing, caching, and communication, have mostly evolved separately, recent needs to support new emerging applications have pushed for an increasing level of integrated design. Two research fields, i.e., content distribution and distributed computing (or processing networks) have focused on the integration of caching-communication and computing-communication technologies into network design, respectively. However, related studies have considered relatively simplified network and service models. For example, mobile edge computing (MEC) studies mainly focus on single task offloading, and service function chaining (SFC) on single processing pipelines (as illustrated in Fig. \ref{fig:CCC_network}), resulting in optimization and control techniques that exhibit sub-optimal performance when applied to the generic Metaverse scenario.

The main advantages of a truly integrated 3C infrastructure include: (i) the {\em joint} optimization of 3C technologies enables higher operational efficiency and QoE (quality of experience), (ii) the adaptability of 3C resource allocations makes the network more flexible and resilient to changing network conditions and service demands, and (iii) the scale and heterogeneity of 3C-equipped network nodes increase service availability and opportunities for enhanced performance. 

Maximizing the benefit of this promising paradigm rests on the design of new tools and methods for the end-to-end optimization and dynamic control of such complex global infrastructure. For example, when a service request emerges, the network control policy needs to {\em coordinate} the selection of
(i) caching locations to provide digital objects,
(ii) computation locations to execute service functions, and
(iii) communication paths to route all associated data streams,
{\em jointly} optimized with dynamic decisions on (iv) traffic scheduling and (v) resource allocation at all network locations.

\section{Cloud Network Flow Framework}

In this section, we outline a comprehensive mathematical framework developed over the past few years, we term {\em cloud network flow}, that enables the end-to-end optimization and dynamic control of completely general Metaverse applications over 3C NextG networks \cite{cai2022delay,cai2022multicast_arxiv,cai2022CCC_arxiv,cai2022xpipelines,cai2021multicast,cai2021delay,cai2020mec,poularakis2020placement,FenLloTulMol:J18a,FenLloTulMol:J18b,Feng_et_al_2017_Asilomar,Feng_et_al_2017_ICC,Feng_et_al_2017_Infocom,Feng_Molisch_2016_TIT,barcelo2016IoT,Feng_et_al_2016_ICC,Feng_et_al_2016_INFOCOM}. While the description here is kept in high-level form, we refer the reader to the above cited papers for further details, analysis, and results.

\subsection{General System Model}

\subsubsection{Service DAG Model \cite{cai2022CCC_arxiv}}

\begin{figure}[t]
    \centering
    \includegraphics[width = \columnwidth]{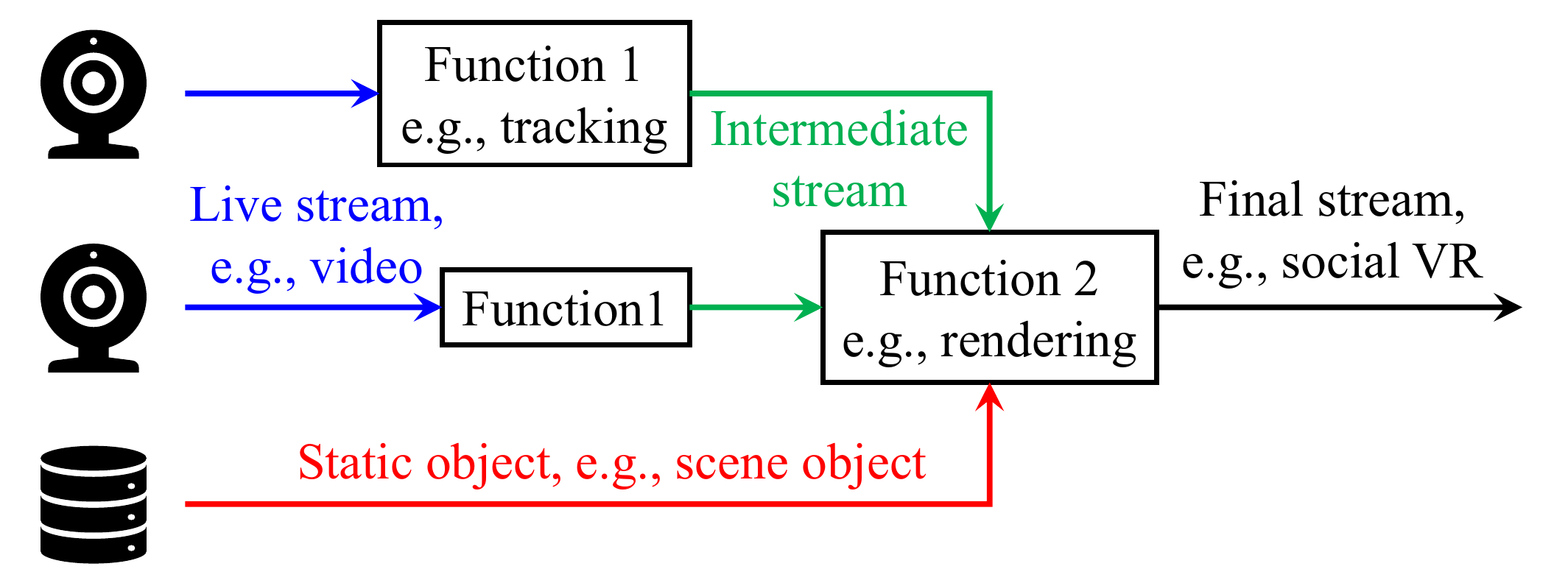}
    \caption{
    Service DAG model for general Metaverse applications.
    }
    \label{fig:service}
\end{figure}

First, we introduce a generic Metaverse application model, referred to as {\em service directed acyclic graph (DAG)}, used to describe the set of service functions, and the source and intermediate data streams they operate on, required to generate the final experience, as shown in Fig. \ref{fig:service}. Each function is characterized by three parameters: {\em merging ratio}, {\em workload}, and {\em scaling factor}, describing the ratio of individual input stream sizes, the computational resource consumption, and the generated output stream size, per unit of input data, respectively. In general, the input streams include {\em live data} (collected by sensors) and {\em static objects} (pre-stored in the network), with associated network locations referred to as {\em live sources} and {\em static sources}, respectively \cite{poularakis2020approximation}. We note that a static object can be provisioned (via replication) by {\em any} of the associated static sources (i.e., caching locations) in an {\em on-demand} manner (per service function's request).

\subsubsection{NextG Network Model}

Next, we describe a practical NextG network model that allows characterizing highly heterogeneous and dynamic systems (as shown in Fig. \ref{fig:CCC_network}) with non-uniform resource distribution in both space and time dimensions. In particular, the processing/transmission capacity of any node/link can be modeled by $C(t) = C( \omega(t), \alpha(t) )$, impacted by uncontrollable system states $\omega(t)$ (e.g., channel state), and controllable resource allocation decisions $\alpha(t)$ (e.g., transmitted power).

\subsection{Policy Design}

The cloud network flow framework is established for control policy design in 3C networks, encompassing network-layer packet processing, transmission, and replication operations, as well as physical-layer multi-dimensional resource allocation decisions.

\subsubsection{Cloud Network Flow}

\begin{figure}[t]
    \centering
    \includegraphics[width = \columnwidth]{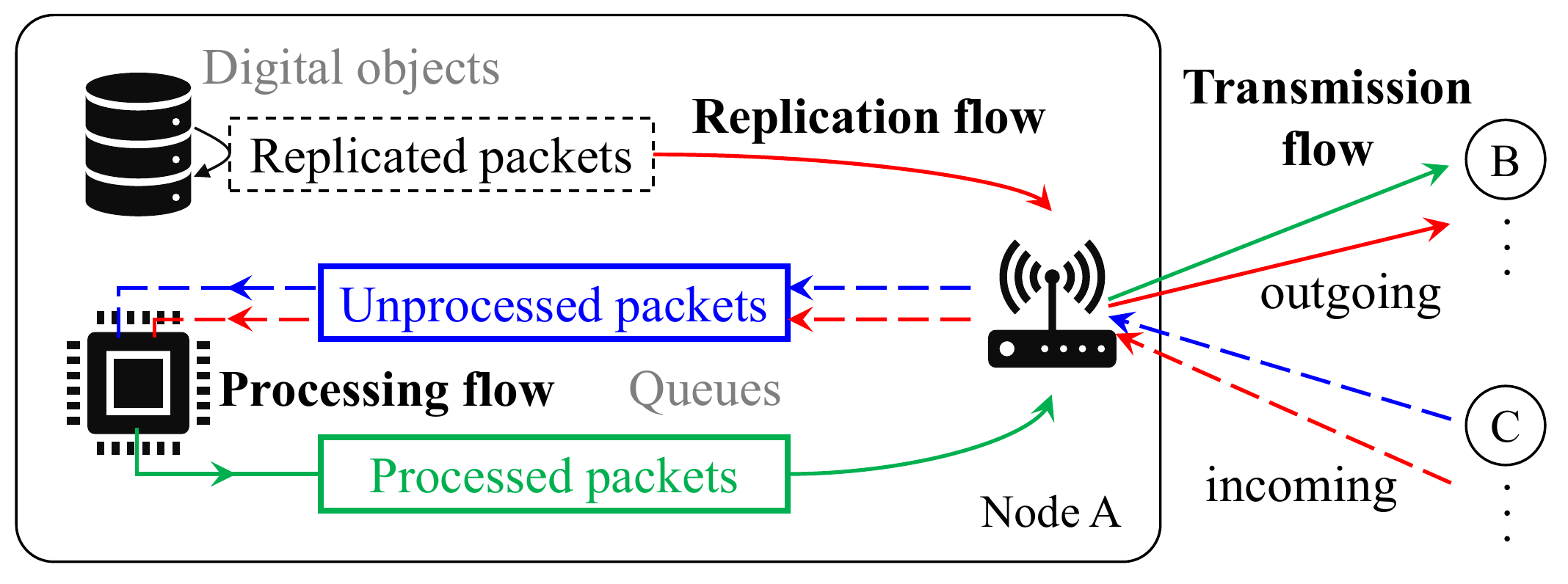}
    \caption{Cloud network flow model.}
    \label{fig:flow}
\end{figure}

We use cloud network flow variable $f(t)$ to represent dynamic control {\em actions} taken on the 3C infrastructure, i.e., the amount of data packets scheduled for 3C operations. As illustrated in Fig. \ref{fig:flow}, the cloud network flow model includes transmission flows (characterizing packet forwarding between neighbor nodes), processing flows (characterizing packet processing via local service functions), and replication flows (characterizing the creation of packet replicas from static objects) \cite{FenLloTulMol:J18a,cai2021multicast,cai2022CCC_arxiv}.

There are two main constraints imposed on the flow variables:
(i) {\em Service chaining} constraints, which impose the relationship between input and output flows as they undergo service function processing (including merging ratio and scaling factor parameters); and
(ii) {\em Capacity} constraints, which limit resource usage: the incurred resource consumption shall not exceed the allocated capacity $C(t)$, via which physical- and network-layer decisions interact.

\subsubsection{Queuing System}

Cloud network flow employs a flexible and generalized queuing system to characterize the evolution of the network {\em state} resulting from the resource allocation and flow scheduling actions taken on the 3C infrastructure. The cloud network flow queuing system $Q(t)$ can include physical queues representing the build-up of packets (i) at different stages of a service DAG (to drive packet processing operations) \cite{FenLloTulMol:J18a,FenLloTulMol:J18b,cai2020mec}, (ii) with different lifetimes (to drive packet dropping and scheduling operations under strict latency constraints) \cite{cai2021delay,cai2022delay}, and (iii) with different replicating status (to drive packet replication operations for multicast services) \cite{cai2021multicast,cai2022multicast_arxiv}.

In addition, the framework is flexible enough to admit the definition of virtual queues that allow characterizing other relevant metrics such as anticipated resource loads (exploiting global knowledge) \cite{zhang2021multicast} and destination-driven computation and data demands \cite{yeh2021deco}.

\subsubsection{Formulation}

We define the system state $s(t)$ and action $a(t)$ to aggregate corresponding physical- and network-layer quantities, given by: 
\begin{align}
    s(t) = (\omega(t), Q(t)),\ a(t) = (\alpha(t), f(t)),
\end{align}
based on which QoE metrics (e.g., throughput, latency) can be expressed as
\begin{align}
    E_i(t) = E_i(s(t), a(t)),\ i \geq 0.
\end{align}

We then formulate the following sequential decision making problem over $\{ a(t): t\geq 0 \}$:
\begin{subequations} \label{eq:MDP} \begin{align}
    & \max \hspace{5pt} \overline{E_0( s(t), a(t) )} \\ 
    & \st \hspace{8pt} \text{QoE constraints } \overline{E_i( s(t), a(t) )},\ i\geq 1, \\
    & \hspace{30pt} \text{Chaining and Capacity constraints} \label{eq:service}
\end{align} \end{subequations}
involving multiple QoE metrics, where $\overline{E_i(t)}$ denotes the average performance in terms of metric $E_i(t)$. Under reasonable assumptions, \eqref{eq:MDP} can be transformed into a (constrained) Markov decision process (MDP) problem.

\subsubsection{Solutions}

The formulated (MDP) problem admits standard solutions, e.g., reinforcement learning (RL) \cite{mnih2015human}. However, while RL methods yield effective solutions for single-agent learning problems, the multi-agent RL variants \cite{lowe2017multi} required to address the end-to-end control of large-scale 3C networks for Metaverse applications can lead to inefficient and unstable training procedures (without convergence guarantees) that result in sub-optimal performance.

A special case of the network control problem \eqref{eq:MDP} that has important practical relevance is the setting in which we only have an action-dependent objective and stability constraints \cite{Nee:B10}, i.e.,
\begin{subequations} \label{eq:LDP} \begin{align}
    & \max \hspace{5pt} \overline{E_0( a(t) )} \\ 
    & \st \hspace{8pt} \overline{E_1( s(t), a(t) )} = \overline{Q(t)} < \infty, \\
    & \hspace{30pt} \text{Chaining and Capacity constraints.}
\end{align} \end{subequations}
This problem can be solved leveraging Lyapunov drift-plus-penalty (LDP) control \cite{Nee:B10}, guiding the design of two important classes of network control policies with insightful interpretations: (i) Distributed policies, e.g., DCNC \cite{FenLloTulMol:J18a,Feng_et_al_2016_ICC,Feng_et_al_2016_INFOCOM}, DWCNC \cite{FenLloTulMol:J18b,Feng_et_al_2017_Asilomar,Feng_et_al_2017_ICC}, MECNC \cite{cai2020mec}, DECO \cite{yeh2021deco}, where routes are dynamically adjusted based on local queue observations, e.g., {\em differential backlog} between neighbor nodes. (ii) Centralized policies, e.g., UMW \cite{SinMod:J18}, UCNC \cite{zhang2021multicast}, DI-DCNC \cite{cai2022CCC_arxiv}, where routes for each incoming packet are selected based on global queuing states. These LDP based solutions can achieve optimal throughput and, for some of them, operational cost, while guaranteeing bounded average delay, without the need of expensive training (required for standard MDP solutions).

\subsection{Advanced Solutions and Open Problems}

\subsubsection{End-to-End Latency}
\label{sec:latency}

Timely content delivery is crucial to interactive Metaverse applications. To this end, our recent works \cite{cai2021delay,cai2022delay} address the challenging problem of {\em timely throughput} analysis and optimization, giving rise to a control policy that makes {\em lifetime} driven routing and scheduling decisions.\footnote{
    In this study, each packet is assigned a strict deadline. The packet lifetime indicates the {\em time to live} (or {\em time till deadline}), and the
    timely throughput the {\em on time} (or {\em by deadline}) {\em service delivery rate}.
}
When dealing with general Metaverse applications involving multiple concurrent pipelines, 
the end-to-end service delay should be taken as the {\em maximum} over the concurrent pipelines, requiring new extensions to the problem formulation and associated optimization methods.

\subsubsection{Multiple Destinations}

The social and interactive nature of Metaverse applications results in many data streams being shared and simultaneously consumed by multiple users/destinations. In our recent works \cite{cai2021multicast,zhang2021multicast,cai2022multicast_arxiv}, we have developed multicast cloud network control policies that leverage {\em in-network packet replication} to enhance multicast content distribution. Another promising way to achieve this goal is by exploiting the {\em broadcast} nature of the wireless medium -- incorporating it into the current framework is object of further investigation.

\subsubsection{Multiple Pipelines}

To jointly handle the multiple service pipelines involved in Metaverse applications, our recent work \cite{cai2022CCC_arxiv} addresses the problem of {\em coordinated multiple stream routing}, including a control policy that guarantees packet routes from different pipelines to meet at common processing locations. Nonetheless, the problem of {\em coordinated multiple stream scheduling} remains unsolved and of interest for future work.

\section{Example Results}

In this section, we conduct numerical experiments to demonstrate the benefit of the envisioned 3C infrastructure and associated control framework in the context of a mobile VR application. The VR application generates the consumed 3D field-of-view (FOV) from digital 2D FOV images selected based on the user's tracking information, as shown in Fig. \ref{fig:requirement}. We employ the algorithm in \cite{cai2022CCC_arxiv} to perform joint dynamic decisions on static source (i.e., cache) selection, processing location, and route selection for each service request.

\subsection{Experiment Setup}

Consider a $100\,\text{m}\times 100\,\text{m}$ square area with $100$ randomly moving users and a base station (BS) at the center. Each user requests the described VR application at a constant rate $\lambda$ (i.e., refresh rate, in {\em frames per second}, fps), and the requested 2D images are selected from a library (of $10^4$ images) following a Zipf distribution of parameter $\gamma_{\text{p}} = 1$. Each 2D image has a size of $3$ Mb, and the VR processing requires $3\times 10^7$ computing cycles to generate a $6$ Mb 3D FOV \cite{sun2019mobileVR}.

\begin{figure}[t]
    \centering
    \includegraphics[width = \columnwidth]{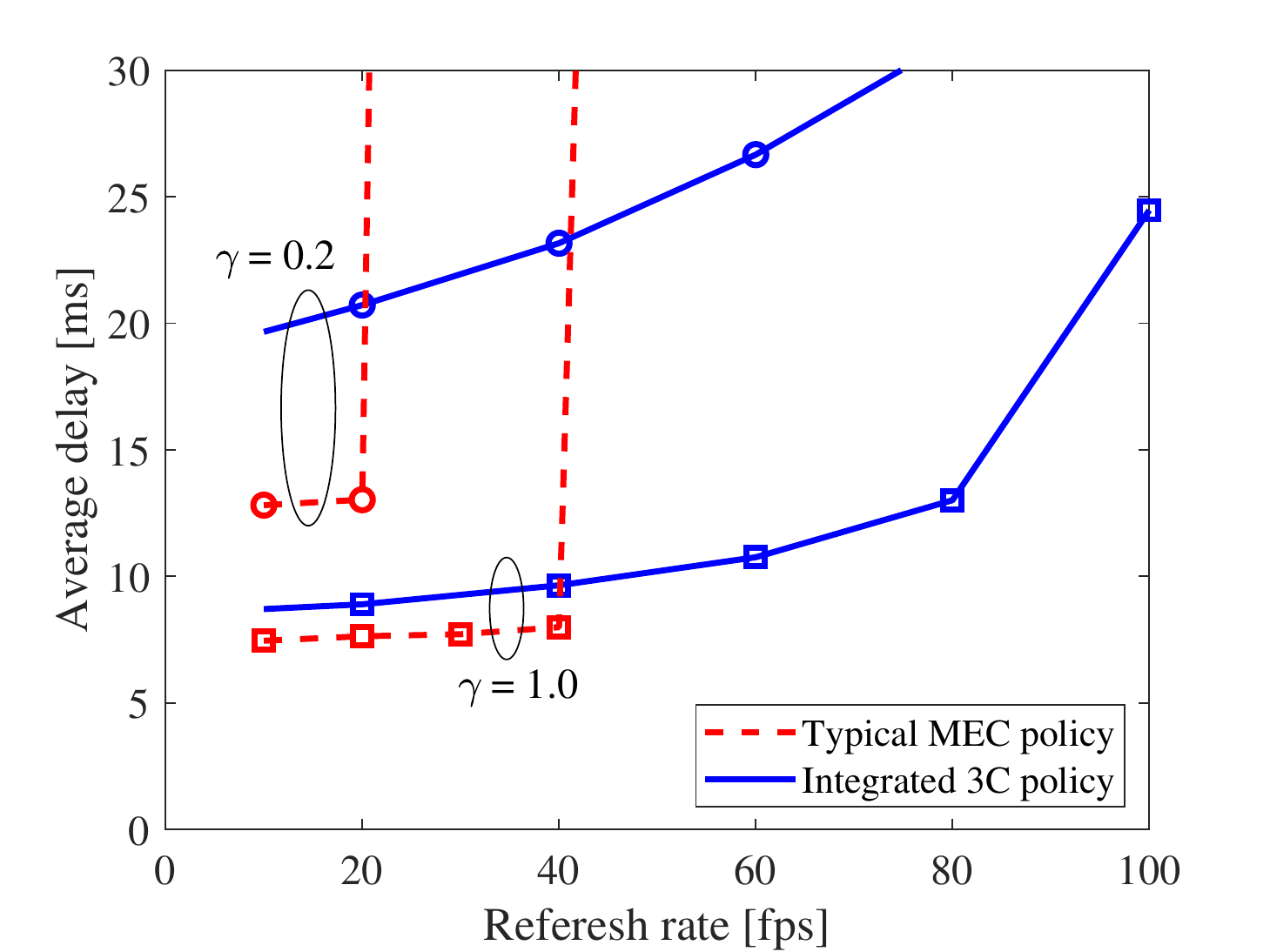}
    \vspace{-10pt}
    \caption{Effects of 3C integration (storage capacity $\beta_3 = 30\%$).}
    \label{fig:policy}
\end{figure}

Each user is equipped with a $3$ GHz processor, and can cache part (a ratio of $\beta_3$) of the library according to some caching distribution (assumed to be another Zipf distribution of parameter $\gamma$). The BS is equipped with $10$ identical processors and has the entire library stored. Users can collaborate via device-to-device (D2D) communication: at every time slot, each user can communicate with one of its neighbors within a cooperation range of $20$ m, with $1$ W transmission power and $20$ MHz bandwidth. The BS can select $20$ users to serve in each slot, at a rate of $200$ Mbps for each of them.

\subsection{Results}

\subsubsection{Impact of 3C Integration}

We first compare the performance of the 3C network running the developed control policy with respect to a typical MEC scenario focusing on individual users offloading tasks to the BS. The users' storage capacity is set to $\beta_3 = 30\%$, and two caching distributions, $\gamma = 0.2$ and $1$, are evaluated.

The results are shown in Fig. \ref{fig:policy}, and we make three observations. First, the integrated 3C policy effectively improves the attained throughput (the blue-square curve achieves $90$ fps under $20$ ms delay requirement), critical for higher VR QoE. Second, the caching distribution significantly impacts the throughput and delay performance, as different $\gamma$ values lead to different average hop distances to find the requested images. Finally, the delay attained by the algorithm in the low-congestion regime (e.g., $\lambda \leq 40$ fps) is not ideal, which can be mitigated by the modified design described in Section \ref{sec:latency}.

\subsubsection{Multi-Dimensional Resource Tradeoffs}

\begin{figure}[t]
    \centering
    \includegraphics[width = \columnwidth]{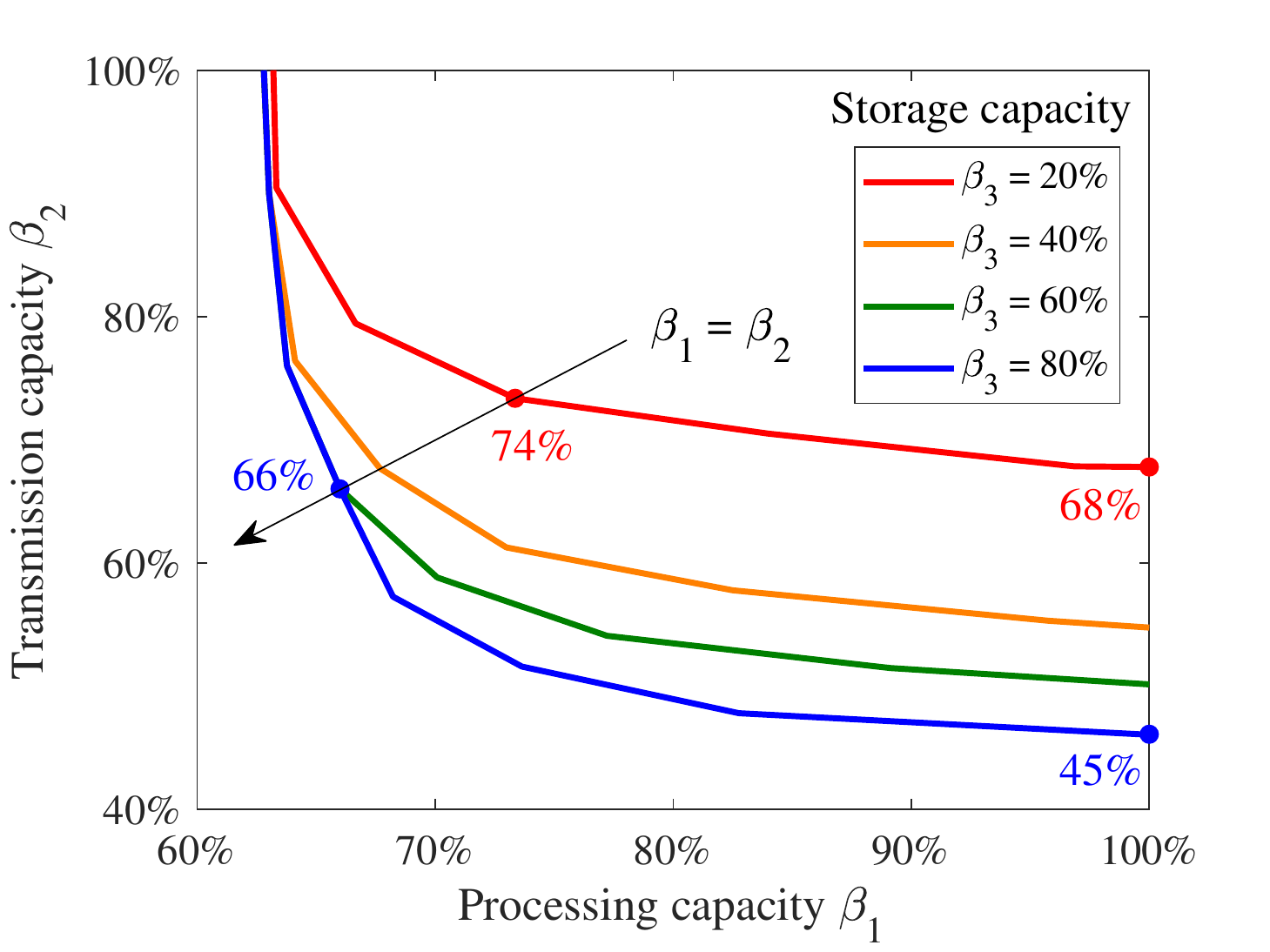}
    \vspace{-10pt}
    \caption{Tradeoffs between multi-dimensional resources ($\gamma = 1$).}
    \label{fig:tradeoff}
\end{figure}

Next, we evaluate the tradeoffs between processing, transmission, and storage resources, assuming $\lambda = 60$ fps refresh rate and $20$ ms delay requirement. The available processing and transmission capacities at each user are given by $\beta_1$ and $\beta_2$ (in percentage of corresponding maximum budgets), respectively. We then define the {\em feasible region} as the collection of $(\beta_1, \beta_2)$ pairs under which the delay requirement is fulfilled.

Fig. \ref{fig:tradeoff} depicts the feasible regions with different storage capacities (using the caching distribution of $\gamma = \gamma_{\text{p}} = 1$). Since higher QoE (e.g., lower latency) can be attained with more resources, i.e., $(\beta_1, \beta_2) \to (1, 1)$, the feasible regions are to the upper-right of the border lines. As we increase the storage resource $\beta_3$ from $20\%$ to $80\%$, more processing and transmission resources can be saved: when $\beta_1 = \beta_2 = \beta_{12}$, the resource saving ratio, i.e., $1 - \beta_{12}$, grows from $26\%$ to $34\%$. A larger saving ratio can be achieved when further narrowing the focus on one resource dimension: when $\beta_1 = 1$, i.e., all devices operate at maximum {\em compute} capacity, even more transmission resources, e.g., {\em bandwidth}, can be saved, i.e., $1 - \beta_2$, from $32\%$ to $55\%$ (provided the same storage resources).

\section{Conclusions}

In this paper, we first illustrated the generality and relevance of Metaverse applications and their unique multi-dimensional requirements. We then described the main characteristics of the envisioned Metaverse-ready compute-, communication-, and data-intensive infrastructure, followed by a cloud network flow mathematical framework for its end-to-end optimization and dynamic control. Numerical results validated the promising role of the envisioned 3C infrastructure to support Metaverse applications, and of the described framework for its efficient operation.


\end{document}